# Optimal Control of Fluid Dosing in Hemorrhage Resuscitation

Jacob Grant
*College of Aeronautics and Engineering*
Kent State University
Kent, Ohio
jgrant29@kent.edu

Hossein Mirinejad
*College of Aeronautics and Engineering*
Kent State University
Kent, Ohio
hmiri@kent.edu

*Abstract*— Fluid resuscitation, also called fluid therapy, is commonly used in treatment of critical care patients. Of the variables inherent to fluid therapy treatment, fluid infusion types are well-studied, but volumetric optimization is not. Automated fluid resuscitation systems employ computational control algorithms along with physiological sensors and actuators to automatically adjust the fluid infusion dosage in hemorrhage scenarios. Most automated hemorrhage resuscitation control algorithms implemented to date are of empiric nature and not effective in the presence of clinical disturbances. This work presents preliminary results of a novel controller based on optimal control approach for automated fluid therapy systems. A receding horizon controller is designed to automatically adjust fluid infusion rate in hemorrhagic situations. The proposed control approach aims to address the inefficiency of current resuscitation control algorithms in finding the accurate fluid dosage adjustments. Performance of the proposed approach is compared against a proportional-integral-derivative (PID) controller for a simulated hemorrhage scenario.

## I. Introduction

Traumatic injuries account for 10% of the global burden of disease and nearly 40% of the mortality associated with them is due to uncontrolled or insufficiently controlled hemorrhage. Fluid resuscitation is vital to effectively control hemorrhage and reduce mortality and morbidity associated with hemorrhagic injuries. Several factors are critical for successful resuscitation outcomes, including the type of fluid and the amount of infusion dosage used in resuscitation [1]. While the type of fluids used in resuscitation has been intensively examined, accurate fluid dose has not been well studied. Administration of proper infusion dosage adjustments to restore cardiac functions and stabilize patients is challenging [2] and often prone to human errors in clinical settings [3-4]. Thus, a closed-loop control algorithm that automatically manages infusion dosage adjustments may eliminate human errors, allow for caregiver attention on other factors, and be widely applied to a large number of patients where the supply of critical medical resources is scarce, such as during a mass casualty incident [5].

Automated fluid management systems incorporate an embedded control algorithm that targets optimizing a hemodynamic endpoint such as blood volume or arterial pressure to automatically adjust the fluid infusion dosage in different scenarios e.g., hemorrhage or hemorrhaging shock [5-10]. In [6], a rule-based control algorithm (decision table) based on the pulse pressure variation was designed for the fluid management of a simulated hemorrhage scenario. In [7], a model-based controller with volumetric feedback built and regulated an individualized control law using a low-order lumped-parameter model. In [8], a fuzzy logic controller was used targeting mean arterial blood pressure (MAP) as the endpoint during three separate bleeds. The integration of a proportional-integral control algorithm, which utilizes the error value in a ratio to the output value to determine subsequent correctional values, was explored in [9]. These control approaches lack the capability of accurately adjusting the infusion dosage and are not effective in the presence of clinical disturbances due to their ad-hoc and empiric nature, as demonstrated in the literature [5, 10].

This paper presents an advanced control algorithm based on optimal control design techniques for automated fluid management. By formulating fluid resuscitation as an optimal control problem, where the target variable to optimize is the patient's blood volume and input variables are both fluid infusion and hemorrhage rate, the RBF-Galerkin optimization method can be applied to efficiently solve the optimal control problem. RBF-Galerkin is a high-accuracy numerical method recently developed by Mirinejad et al [11] for solving constrained optimal control problems. A receding horizon controller (RHC) based on the RBF-Galerkin method is designed for optimized infusion dosage recommendations.

The rest of this paper is organized as follows: Section II addresses the theory and methodology for designing the controller. Section III-A describes the results for a simulated patient without measurement noise, and Section III-B shows the results when noise is simulated. Conclusions are presented in Section IV.

## II. Methodology

A new control algorithm based on the RBF-Galerkin optimal control approach [11] was designed to optimize fluid infusion dosage in hemorrhagic scenarios. The hemorrhage resuscitation was formulated as an optimal control problem, and a receding

This work was supported in part by the Summer Undergraduate Research Experience (SURE) Program at Kent State University, Kent, Ohio.

horizon controller based on the RBF-Galerkin solution was proposed to find fluid dosing at each iteration.

Direct methods are extensively used for solving optimal control problems by applying approximation-then-discretization to the original problem [12-14]. In direct transcription, the states and/or controls are approximated by a specific function that has unknown coefficients and the optimal control problem is discretized using a set of proper points to transcribe it to a nonlinear programming (NLP) problem. This resulting problem is solved by readily-available NLP solvers. The controller outlined in this work uses a highly efficient direct method called the RBF-Galerkin method [11] for solving fluid dosing optimal control problems.

Consider a general optimal control problem, in which the state is shown as $x(\tau) \in \Re^n$, the control as $u(\tau) \in \Re^m$, the initial optimization time as $t_0$, and the final optimization time as $t_f$. The goal is to minimize the cost functional

$$J = \Gamma(x(-1), t_0, x(1), t_f) + \frac{t_f - t_0}{2} \int_{-1}^{1} L(x(\tau), u(\tau)) d\tau \quad (1)$$

subject to state dynamics

$$\dot{x}(\tau) = \frac{t_f - t_0}{2} f(x(\tau), u(\tau)), \quad (2)$$

boundary conditions

$$\gamma(x(-1), t_0, x(1), t_f) = 0 \in \Re^\gamma, \quad (3)$$

and its mixed state-control path constraints

$$q(x(\tau), u(\tau)) \leq 0 \in \Re^\gamma. \quad (4)$$

The time interval $\tau \in [-1,1]$ can be transformed to $t \in [t_0, t_f]$ via

$$t = \frac{(t_f - t_0)}{2} \tau + \frac{t_f + t_0}{2}. \quad (5)$$

**RBF-Galerkin Method:** The radial basis function (RBF) is defined as a real-valued function with a fixed center point, whose output value depends on the distance from the center point, i.e.,

$$\rho(y, c) = \rho(||y - c||) \quad (6)$$

where $\rho$ is the RBF, $c$ is the center point, and $|| \: ||$ is the Euclidean norm. In general, an RBF could be piecewise smooth like Polyharmonic Splines or infinitely smooth such as Gaussian RBFs or Multiquadric RBFs. Infinitely smooth RBFs are sometimes called global RBFs. The RBF-Galerkin method [11] uses global RBFs for approximating the states and controls of the optimal control problem, applies Galerkin error projection to residuals –the difference between the RBF approximated constraints and the actual constraints–, and incorporates an arbitrary discretization scheme to transform the optimal control problem to the NLP problem (see [11] for more information). The hemorrhage resuscitation model of [15] was used relating blood volume response to fluid infusion dosage as

$$\ddot{\tilde{V}}_B(t) + K\dot{\tilde{V}}_B(t) = \frac{[\dot{u}(t) - \dot{v}(t)]}{V_{B0}} + \frac{K[u(t) - v(t)]}{V_{B0}(1+\alpha)} \quad (7)$$

where $u$ is the fluid infusion, $v$ is the hemorrhage rate, $K$ is the rate of fluid shift between intravascular and interstitial compartments, $\alpha$ is a constant parameter value which varies according to the patient's overall physiological state, and $V_{B0}$ is the initial blood volume. Also, $\tilde{V}_B(t)$ represents the normalized value of change of blood volume, i.e., $\tilde{V}_B(t) = \triangle V_B(t)/V_{B0}$, where $\triangle V_B(t)$ is the change of blood volume.

The first step to designing the control algorithm consisted of formulating the hemorrhage resuscitation optimal control problem. To this end, patient parameters from [15] were used, and the cost functional of (1) was set to minimize the difference between the actual $\tilde{V}_B(t)$ and target $\tilde{V}_B(t)$. The maximum permissible infusion volume was set to 25 mL/kg. Using this information and model of (7), the optimal control problem was formulated and the RBF-Galerkin method was employed to transcribe it to an NLP problem solved by SNOPT solver [16].

The RBF-Galerkin's flexibility in optimal control discretization and its high accuracy make it ideal for addressing the disturbances present in clinical fluid resuscitation scenarios, discussed further in Section III. Once the optimal control problem was deemed effective, the fluid management controller was designed in an RHC scheme. RHC, also known as model predictive control (MPC), is a general-purpose control scheme that involves repeatedly solving a constrained optimization problem, using predictions of future costs, disturbances, and constraints over a moving time horizon to choose the control action [17]. The optimal infusion dose sequence for the entire simulation was computed by the RBF-Galerkin method from the current state to the desired state over a finite time horizon. However, only the first dose of the sequence produced was given to the patient, and the state was updated by measuring the patient's current blood volume level using the model of (7). The finite horizon optimization problem was repeated using the updated state and the recent control as the initial values for the optimal control problem. The resulting control approach is called the RBF-Galerkin based RHC method for fluid dose recommendations.

## III. RESULTS & DISCUSSION

The proposed controller was designed in MATLAB for several resuscitation cases. The preliminary results for a simulated case are demonstrated here. This scenario was incorporated from [18], where a moderate hypovolemia was applied to volunteer human subjects by withdrawal of 900 mL blood prior to resuscitation (no active hemorrhage used due to ethical reasons). In this scenario, the baseline and target blood volume were set to 3,940 mL and 5,000 mL, respectively. The infusion dosage was limited to be between 0 and 25 mL/kg, a maximum dosage derived from [8]. The simulation was then run for a sixty minute interval, i.e., the algorithm iterated sixty solutions of the control problem. Performance of the proposed control algorithm was compared against a conventional proportional-integral-derivative (PID) controller simulated in MATLAB. The PID gains were tuned using MATLAB PID Tuner app and the results are illustrated in the same figures.

*A. Blood Volume Measurement without Observational Error*

Fig. 1. demonstrates the achieved blood volume levels along with the fluid dose adjustments from the RBF-Galerkin based RHC for nominal values of parameters (without measurement error).

The proposed controller increases the blood volume to the target level within the optimization time, given the conditional parameters and the cost functional. As illustrated in Fig. 1a, a noiseless environment produces a smooth transition from initial to target volume. The infusion rate, illustrated in Fig. 1b, can be seen to spike at the halfway point before decreasing, reflected in the slope of the blood volume graph. This spike is indicative of the algorithm's several assessments: that reaching target volume as quickly as possible without over-infusion requires climbing infusion rates (from 2-30 minutes), and gradual target volume achievement afterwards requires a subsequent decrease in rate to safely transition from speed to accuracy. Compared to the PID controller, the RHC was volumetrically more efficient and avoided unnecessarily high infusion dosage in the beginning of treatment (see Fig 1b). the optimal infusion rate much sooner. Various local minima and maxima illustrate minor adjustments by the algorithm to perfectly achieve the target value at peak efficiency.

*B. Blood Volume Measurement with Observational Error*

A white noise with maximum amplitude of ±250 mL was added to the blood volume to simulate clinical disturbances up to 5%. This is a clinically relevant assumption for blood volume measured by gravimetric or bag calibration technique, hemoglobin concentration technique, or dye dilution technique, according to [19]. Fig. 2 demonstrates the achieved blood volume and fluid dose recommendations by the RBF-Galerkin based RHC in the presence of measurement noise.

As clearly seen in Fig. 2a, the noise causes fluctuations in the measured blood volume. The algorithm is still capable of achieving target volume, and does so with a similar infusion rate as evidenced by Fig. 2b. Note that, even with noise, the RHC remained more efficient in achieving target blood volume and gradually changed the infusion dosage throughout the simulation to meet the BV target. Conversely, the PID controller aggressively started the resuscitation with the maximum dosage of 25 ml/Kg, which did not seem to be efficient or necessary. It should also be noted that these graphs and data are only representative of one scenario, simulated with one virtual patient; other initial and goal parameters could easily be considered within the algorithm for different scenarios.

The results indicate the control algorithm's efficacy in achieving a target value given a series of patient parameters. Adjusting the K or α values from patient to patient proves trivial within the syntax of the controller, and other parameters (such as hemorrhage rate) could also be adjusted as necessary to either simulate scenarios or react and control a real-world scenario.

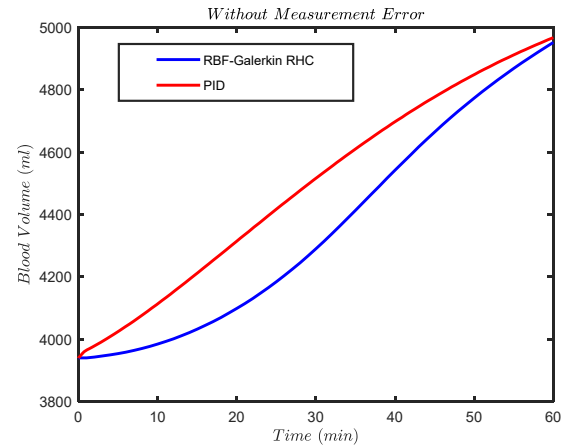

(a)

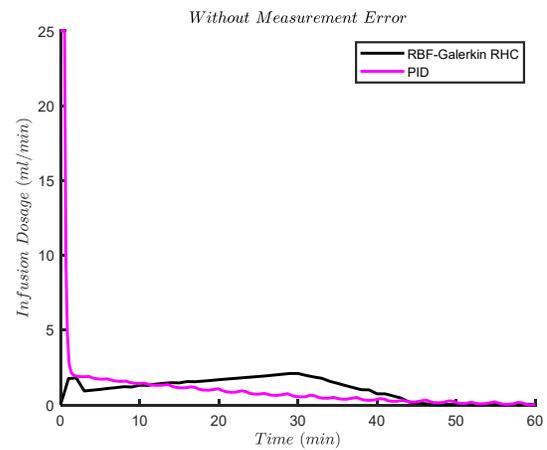

(b)

Fig. 1. (a) Achieved blood volume levels and (b) Fluid dose adjustments from the RBF-Galerkin based RHC and PID controller for blood volume measurements without error

MATLAB simulations were capable of easily and efficiently handling the computational requirements of the code. Perhaps most importantly, the miniscule adjustments by the algorithm to adjust infusion dosage over time show that volumetric optimization can be achieved to a high degree of accuracy and efficiency – far beyond that of commonly employed rule-based controllers or techniques predominantly overseen by human assessment.

It is important to note that while blood volume was the endpoint variable in this formulation of the fluid resuscitation problem, other hemodynamic endpoints such as mean arterial pressure or cardiac output can be used in the proposed strategy. In addition, the proposed strategy is not limited to the current model. In fact, it is worth considering other hemodynamic models such as [7], which could subsequently be compared to the model employed during this research. Furthermore, a future literary review that identifies scenarios of active hemorrhage with specific numerical data on fluid loss volume and rate over time will enable a simulation using this controller. Subsequent comparisons will allow for further assessment of controller response to more extreme disturbances and complications in clinical settings.

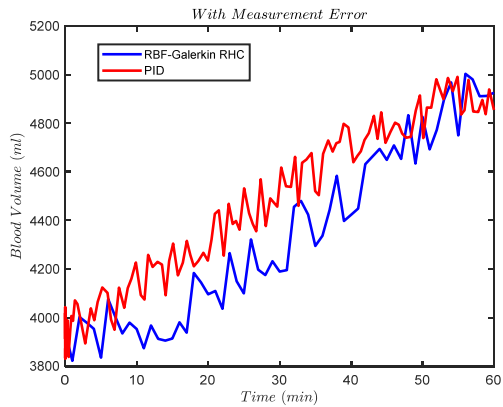

(a)

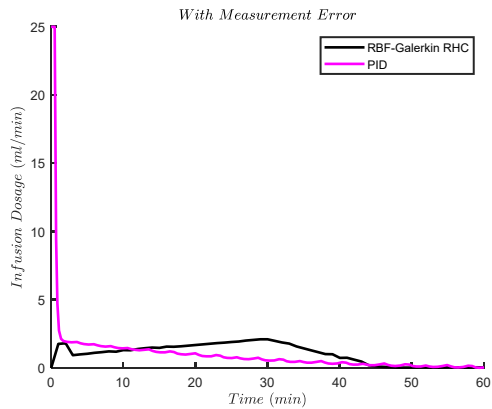

(b)

Fig. 2. (a) Achieved blood volume and (b) Fluid dose adjustments from the RBF-Galerkin based RHC and PID controller in the presence of blood volume measurement error

## IV. Conclusion

Formulating the fluid resuscitation as an optimal control problem and finding the optimal infusion dosage with the RBF-Galerkin method established a viable foundational code and proved the efficacy of a controller developed for specific integration with the RBF-Galerkin method. An RBF-Galerkin based RHC controller was designed to optimize fluid dosing in hypovolemic scenarios. Implementing measurement noises to the data illustrated the controller's capability of adjusting to noise and achieving the design parameter regardless of reasonable interference. Comparison to a PID controller run under the same conditions clearly demonstrated the RHC's superiority in volumetric efficiency, speed, and accuracy.

While the preliminary results are promising, more rigorous assessment of the proposed approach in various clinical scenarios will be needed. In addition, further comparison studies against other control approaches will be conducted to thoroughly investigate the method against the state-of-the-art. Eventual implementation of the controller would utilize a user-friendly interface for adjusting patient parameters without adjusting the source code. Further development of this system will be undertaken after integration with a hardware-in-the-loop test bed system, as in [5,20], to fine-tune any syntax within the controller and optimize the system for real-world application.